\documentclass[twocolumn,pre,showpacs,eqsecnum]{revtex4}

\usepackage{dcolumn}
\usepackage{amsfonts}
\usepackage{amsmath}
\usepackage{amssymb}
\usepackage{bm}
\usepackage{epsfig}

\DeclareMathOperator{\tr}{tr}

\begin{document}


\newcommand{\brm}[1]{\bm{{\rm #1}}}
\newcommand{\tens}[1]{\underline{\underline{#1}}}
\newcommand{\mm}{\overset{\leftrightarrow}{m}}
\newcommand{\xv}{\bm{{\rm x}}}
\newcommand{\Rv}{\bm{{\rm R}}}
\newcommand{\uv}{\bm{{\rm u}}}
\newcommand{\nv}{\bm{{\rm n}}}
\newcommand{\Nv}{\bm{{\rm N}}}
\newcommand{\ev}{\bm{{\rm e}}}

\title{Smectic elastomer membranes}

\author{Olaf Stenull}
\affiliation{Fachbereich Physik, Universit\"{a}t Duisburg-Essen, Campus Duisburg, 47048 Duisburg,
Germany }

\vspace{10mm}
\date{\today}

\begin{abstract}
We present a model for smectic elastomer membranes which includes elastic and liquid crystalline degrees of freedom. Based on our model, we determined the qualitative phase diagram of a smectic elastomer membrane using mean-field theory. This phase diagram is found to comprise five phases, viz.\ smectic-$A$--flat, smectic-$A$--crumpled, smectic-$C$--flat, smectic-$C$--crumpled and smectic-$C$--tubule, where in the latter phase, the membrane is flat in the direction of mesogenic tilt and crumpled in the perpendicular direction. The transitions between adjacent phases are second order phase transitions. We study in some detail the elasticity of the smectic-$C$--flat and the smectic-$C$--tubule phases which are associated with a spontaneous breaking of in-plane rotational symmetry. As a consequence of the Goldstone theorem, these phases exhibit soft elasticity characterized by the vanishing of in-plane shear moduli. 
\end{abstract}


\maketitle

\section{introduction}
\begin{figure}
\centerline{\includegraphics[width=8.4cm]{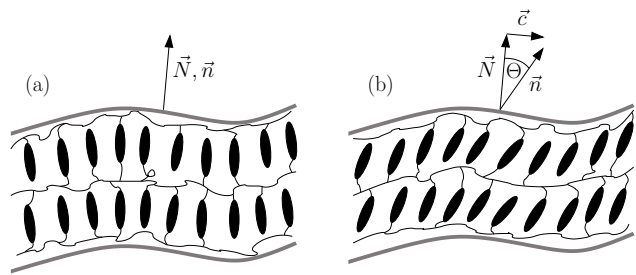}}
\caption{Cartoon of the cross-section of a (a) Sm$A$ and (b) Sm$C$ elastomer membrane. The membranes consist of a few smectic layers such that their height can be neglected in comparison to their lateral extension.}
\label{fig:membraneCartoon}
\end{figure}

Liquid crystal elastomers~\cite{WarnerTer2003} have attracted much attention in recent years because they uniquely combine the rubber elasticity of polymer networks with the anisotropic properties of liquid crystals~\cite{deGennesProst93_Chandrasekhar92} and, therefore, provide exciting challenges for fundamental research (experimental and theoretical) and open new possibilities for novel device applications, for example in sensors and actuators. Essentially, any phase known from conventional liquid crystals can be made in elastomeric form, such as, e.g., nematic, smectic-$A$ (Sm$A$) and smectic-$C$ (Sm$C$). Among the various phases, nematic elastomers have been studied most extensively to date. This research brought about considerable insight into the static and dynamic elastic properties of nematic elastomers, see e.g. Refs.~\cite{golubovic_lubensky_89,FinKun97,VerWar96,War99,LubenskyXin2002,stenull_lubensky_epl2003,Xing_Radz_03,stenull_lubensky_anomalousNE_2003,terentjev&Co_NEhydrodyn,stenull_lubensky_2004,stenull_lubensky_comment}. Smectic elastomers are, at least from the theoretical standpoint, considerably less well understood than nematics. However, there exists substantial literature on their synthesis and their experimental properties, see for example~\cite{shibaev_81_82,fischer_95,bremer&Co_1993,benne&Co_1994,hiraoka&CO_2001,hiraoka&CO_2005,GebhardZen1998,ZentelBre2000} and references below. Usually, smectic elastomers are synthesized by crosslinking side-chain or main-chain liquid crystal polymers. The elaborate crosslinking techniques available to date produce elastic networks that stabilize the liquid crystalline order so that monodomain or single crystal samples result and they leave at the same time sufficient mobility for the mesogenic component to reorient, e.g., when mechanical or electrical fields are applied. With these methods, one can very efficiently synthesize from small amounts of material experimental samples in the form of free-standing thin films or membranes, see Fig.~\ref{fig:membraneCartoon}. Such films have been produced as thin as 75nm~\cite{lehmann&Co_01}, which corresponds to a thickness of about 15 smectic layers, given that the average layer thickness is roughly 5nm. Experiments on such films include measurements of the electroclinic effect in planar or flat samples~\cite{lehmann&Co_01,KohlerZen2005} and measurements of elastic constants of smectic elastomer balloons~\cite{schuering&Co_2001,StannariusZen2002}; samples where smectic elastomer membranes have been inflated to spherical bubbles similar to the inflation of soap bubbles from flat soap films.

Smectic elastomers membranes are possible realizations of anisotropic membranes~\cite{radzihovsky_membraneReview_2004}; a class of systems that has been studied intensively in recent years. Radzihovsky and Toner~\cite{radzihovsky_toner_anisoMem} discovered that permanent in-plane anisotropy qualitatively modifies the phase diagram of polymerized membranes in that it leads to intermediate tubule phases between the usual flat and crumpled phases. Such a tubule phase is a hybrid between the flat and the crumpled phases; the membrane is flat in one direction and crumpled in another. More recently, Xing {\em et al}.~\cite{Xing&Co_fluctNemMem_2003} and Xing and Radzihovsky~\cite{Xing_Radzihovsky_nemTubule_2005} studied nematic elastomer membranes in which in-plane anisotropy is spontaneous rather than permanent. These membranes were shown to have a rich phase diagram comprising isotropic--flat, isotropic--crumpled, nematic--flat, nematic--crumpled and nematic--tubule phases. Because of spontaneous breaking of in-plane isotropy, the nematic--flat and the nematic--tubule phases exhibit a soft elasticity that is qualitatively distinct from the elasticity of the flat and tubule phases of permanently anisotropic membranes.

In this paper, we theoretically study idealized smectic elastomer membranes. Our idealizations, adopted for simplicity, are as follows. First, we assume that the thin films consist of only a few smectic layers such that their height can be neglected in comparison to their lateral extension, i.e., that the membranes can be described as two-dimensional manifolds in three-dimensional space. Second, we entirely neglect self-avoidance, i.e., our model membranes are so-called phantom membranes. Third, we leave aside heterogeneities such as random stresses with must be present in any amorphous solid on grounds of mechanical stability. Moreover, we focus on membranes crosslinked in the Sm$A$ phase that can exhibit, like nematic elastomer membranes, spontaneous in-plane anisotropy and, therefore, by virtue of the Goldstone theorem, can have modes whose energy vanishes with wavenumber. In bulk nematic elastomers and the corresponding membranes and in bulk Sm$C$ elastomers, these Goldstone modes lead to the fascinating phenomenon of soft elasticity whereby, in an idealized limit, certain elastic moduli vanish and thus certain deformations are free of restoring forces. By assuming crosslinking in the Sm$A$ phase, we can expect to find these unusual properties also in the membranes under consideration here.

In the following, we develop a theory for smectic elastomers. This theory has some similarities to the theory for nematic elastomers membranes presented in Ref.~\cite{Xing_Radzihovsky_nemTubule_2005} but it also has considerable differences related to the fact that the isotropic--to--nematic phase transition is generically a first order transition whereas the Sm$A$--to--Sm$C$ transition is generically continuous. As a result, the phase diagrams of nematic and smectic elastomer membranes have qualitatively different topologies. We find that the mean-field phase diagram for a smectic elastomer membranes features Sm$A$--flat, Sm$A$--crumpled, Sm$C$--flat, Sm$C$--crumpled and Sm$C$--tubule phases. The phase transitions between the phases are second order transitions. Among the elasticity of the various phases, that of the Sm$C$--flat and the Sm$C$--tubule phases is most interesting because in these phases, in-plane rotational symmetry  is spontaneously broken and thus, due to the Goldstone theorem, the membrane exhibits soft elasticity. We  investigate the elasticity of these phases in some detail.  

The outline of the remainder of this paper is as follows. Section~\ref{smecticElastomerMembraneModel} presents our model for smectic elastomer membranes. First, some fundamentals of Lagrange elasticity theory are reviewed. Then it is explained how to properly combine elastic and liquid crystalline degrees of freedom to produce a model elastic energy that has the appropriate invariance properties. This elastic energy is presented and its physical contents is explained. Section~\ref{meanFieldPhaseDiagram} analyses our model in mean-field theory to determine the phase diagram qualitatively. Sections~\ref{elasticityOfTheFlatPhase} and \ref{elasticityOfTheTubulePhase}, respectively, treat the elasticity of the Sm$C$--flat and the Sm$C$--tubule phases with emphasis on softness. Section~\ref{concludingRemarks}, finally, contains some concluding remarks.

\section{Model}
\label{smecticElastomerMembraneModel}
Physical membranes are generically two-dimensional manifolds embedded in three-dimensional space. Though it can be worthwhile, e.g.\ when doing field theory, to consider generalizations to $D$-dimensional manifolds in $d$-dimensional space, we will restrict ourselves here for simplicity to the physical case. Generalizations of our model to higher dimension will be straightforward.

To describe smectic elastomer membranes, we need to establish a certain amount of notation. First, let us define what we mean by reference space. This is the space occupied by membrane in its reference confirmation, which we take to be flat. We denote two-dimensional vectors, such as reference space vectors, in bold face and label their components by indices from the beginning of the alphabet, $a, b, c = 1, 2$. We employ the framework of Lagrange elasticity theory. To this end, we label mass points in the undeformed membrane by a reference space vector
\begin{align}
\label{DefIntCoord}
\brm{x} = (x_1, x_2) \equiv (x, y)\, .
\end{align} 
Upon deformation, the membrane assumes some confirmation in the 3-dimensional embedding or target space. We denote target space vectors with arrows and label their coordinates with indices from the middle of the alphabet, $i, j, k = 1, 2, 3$. In particular, we denote the position in target space of the mass point with intrinsic coordinate $\brm{x}$ by
\begin{align}
\label{DefTargetCoord}
\vec{R} (\brm{x}) = (R_1(\brm{x}), R_2 (\brm{x}), R_3 (\brm{x})) \, .
\end{align}
Unless stated otherwise, the summation convention on repeated indices is understood. This applies to the reference and the target space. To keep our discussion as simple as possible, we use orthonormal target space basis vectors $\hat{e}_i$ with components $\hat{e}_{i, j} = \delta_{ij}$ satisfying $\hat{e}_i \cdot \hat{e}_j = \delta_{ij}$ and choose the reference space basis vectors $\hat{e}_a$ to form a subset of the set $\{  \hat{e}_i \}$ as we can because the reference space is a subspace of the target space.

To describe smectic ordering, we employ the unit layer normal $\vec{N} (\brm{x})$ and the Frank director $\vec{n} (\brm{x})$ which describes the local orientation of constituent mesogens. $\vec{n}$ can be decomposed into its components parallel and perpendicular to $\vec{N}$,
\begin{align}
\label{directorDecomp}
\vec{n} = n_\parallel \, \vec{N} + \vec{c} \, ,
\end{align}
where $\vec{N} \cdot \vec{c} = 0$ and $n_\parallel = \sqrt{1 - c_i^2}$, and with $\vec{c}$ being called c-director. To facilitate a discussion of tangent plane vectors such as $\vec{c}$, it is useful to introduce an orthonormal basis of tangent plane vectors $\vec{t}_a$ satisfying
\begin{subequations}
\begin{align}
\vec{t}_a \cdot \vec{t}_b &= \delta_{ab} \, ,
\\
t_{a,i} \, t_{a,j} & = \delta_{ij} - N_i N_j  \, .
\end{align}
\end{subequations}
Any tangent plane vector $\vec{b}$ can be represented in terms of this basis, in which case we denote its components with a tilde, i.e., $\vec{b} = \tilde{b}_a \, \vec{t}_a$. For the c-director, in particular, we have $\vec{c} = \tilde{c}_a \, \vec{t}_a$. Below,  $\tilde{c}_a$ will become a very important quantity and thus we would like to stress here what it stands for physically: $\tilde{c}_a$ represents the components of the c-director in the orthonormal tangent-space basis defined by $\vec{t}_a$.

Now, we will seek an explicit representation of $\vec{t}_a$. Distortions of the reference membrane can be described by the Cauchy deformation tensor $\tens{\Lambda}$~\cite{Love1944,Landau-elas} with components
\begin{align}
\label{lambdaComp}
\Lambda_{ia} = \frac{\partial R_i}{\partial x_a} \equiv \partial_a R_i \, .
\end{align}
The vectors $\vec{T}_a$ defined by $T_{a,i} = \Lambda_{ia}$ lie in the tangent space of the membrane. From these, we can construct the desired orthonormal tangent plane basis vectors via
\begin{align}
\label{tangentPlaneBasis}
t_{a,i} = g_{ab}^{-1/2} \, T_{b,i} \quad \mbox{or} \quad \vec{t}_a = g_{ab}^{-1/2} \, \vec{T}_b \, ,
\end{align} 
where 
\begin{align}
g_{ab} = \vec{T}_a \cdot \vec{T}_b  \quad \mbox{or} \quad \tens{g} = \tens{\Lambda} \, \tens{\Lambda}^T
\end{align} 
is the metric tensor measuring distances in target space between neighboring points, $dR^2 = g_{ab} \, dx_a\, dx_b$. By construction, $\tens{g}$ is invariant under rigid rotations in target space,
\begin{align}
\label{TargetSpaceRot}
R_i \to R_i^\prime = O_{T, ij} \, R_j\, ,
\end{align}
where $\tens{O}_R$ is a target space rotation matrix, and it is positive semi-definite. Note that Eq.~(\ref{tangentPlaneBasis}) implies the relation
\begin{align}
\tilde{b}_a = g_{ab}^{-1/2} \Lambda_{bi}\,  b_i 
\end{align}
between the components of a tangent space vector $\vec{b}$ relative to the bases $\{ \vec{t}_a \}$ and $\{ \hat{e}_i \}$, respectively.
 
Below, we will investigate the various phases of smectic elastomer membranes. The equilibrium confirmations of the membrane (i.e., equilibrium values of its elastic degrees of freedom) in these phases are characterized by certain equilibrium deformation tensors
\begin{align}
\label{lambdaEquiComp}
\Lambda_{ia}^0 =  \partial_a R^0_i \, ,
\end{align} 
or alternatively, up to global rotations in target space, by equilibrium metric tensors
\begin{align}
\tens{g}^0 = \tens{\Lambda}^0 ( \tens{\Lambda}^{0})^T .
\end{align}

Conventionally, elastic energies are formulated in Lagrange elasticity theory in terms of the Cauchy-Saint-Venant~\cite{Love1944,Landau-elas,tomsBook} nonlinear strain tensor 
\begin{align}
\label{defStrainTensor}
\tens{u} = \textstyle{\frac{1}{2}} \big( \tens{g} - \tens{g}^0 \big)  .
\end{align}
This tensor will play an important role further below when we analyze the elastic properties of the soft phases of smectic elastomer membranes.

Having defined various variables describing elastic and liquid-crystalline degrees of freedom, we will now turn to construct a model elastic energy for smectic elastomer membranes. This requires some care, because on one hand, the liquid-crystalline fields $\vec{n}$ and $\vec{N}$ live in the target space, i.e., they transform as (rank 1) tensors in this space and they are scalars in the reference space. The metric tensor and the strain tensor, on the other hand, live in reference space, i.e., they are rank 2 tensors in reference space and scalars in target space. Elastic energies, in general, are invariant under rigid rotations in the target space and under the symmetry transformations of the reference space. Our reference membrane is a flat Sm$A$ one, and thus our model elastic energy has to be rotationally invariant in target and reference space. This requires that we be able to construct combinations of liquid-crystalline fields and reference space elastic variables that are rotationally invariant in both spaces.

Our approach to construct these combinations is based on representing tangent space vectors in terms of the basis $\{ \vec{t}_a \}$. This approach is intimately related to our approach for combining elastic and liquid crystalline degrees of freedom in bulk liquid crystalline elastomers~\cite{stenull_lubensky_letter_2005,stenull_lubensky_SmC,stenull_lubensky_SmA,stenull_lubensky_Sm_dynamics} which is via exploiting the polar decomposition theorem~\cite{HornJoh1991}. To motivate the approach taken here, let us consider the transformation 
\begin{align}
\label{trafoDEF}
x_a \to x_a^\prime = O^{-1}_{R, ab} \, x_b\, ,
\end{align}
where $\tens{O}_{R}$ is a reference space rotation matrix. Under this simple change of basis in reference space, $\Lambda_{ia} \to \Lambda_{ia}^\prime = \Lambda_{ib} \,  O^{-1}_{R, ba}$, and
\begin{align}
\label{trafoGn}
g_{ab}^n \to  O_{R, ac }\, g_{c c^\prime}^n \,  O^{-1}_{R, c^\prime b}\, ,
\end{align}
for any power $n$. For our tangent space basis vectors, this leads to 
\begin{align}
\vec{t}_a \to \vec{t}_a^\prime = \vec{t}_b \,  O^{-1}_{R, ba}\, .
\end{align}
Since a tangent space vector $\vec{b}$ does not change under the simple change of basis (\ref{trafoDEF}), we have $\vec{b}= \tilde{b}_a \, \vec{t}_a = \tilde{b}^\prime_a \, \vec{t}^\prime_a$ with
\begin{align}
\label{trafoV}
\tilde{b}^\prime_a = O_{R, ab} \, \tilde{b}_b  \, .
\end{align}
Equations~(\ref{trafoGn}) and (\ref{trafoV}) imply that combinations of the form $\tilde{b}_a \, g_{ab}^n \, \tilde{b}_b$ are invariant under the transformation~(\ref{trafoDEF}). We will apply this approach to the c-director. This then allows us to construct our model elastic energy from terms of the form $(\vec{N} \cdot \vec{n})^2 = 1 - \tilde{c}_a^2$, $\tilde{c}_a \, g_{ab} \, \tilde{c}_b$ or  $\tilde{c}_a \, u_{ab} \, \tilde{c}_b$ etc., which have the desired invariance properties.

In what follows, we will use the metric tensor, as opposed to the deformation tensor or the strain tensor, as our order parameter field for confirmations. This approach has two advantages: First, it is independent of the actual orientation of the membrane in the target space (as it would if we used the strain tensor). Second, the metric tensor provides for an intuitive distinction between the different phases., i.e., the two eigenvalues of the equilibrium tensor $\tens{g}^0$ encode how much the membrane is extended along the principal axes in reference space. In a flat phase, both eigenvalues are larger than zero. In a crumpled phase, both eigenvalues vanish. In a tubule phase, where the membrane is extended along one principal axis and crumpled along the other, $\tens{g}^0$ has only one non-vanishing eigenvalue. To determine the liquid-crystalline order of our smectic elastomer membrane, we will use as order parameter fields the components $\tilde{c}_a$ of the c-director in the basis $\{ \vec{t}_a \}$. If both equilibrium values $\tilde{c}_a^0$ vanish, the director has no component in the smectic plane and the membrane is in a Sm$A$ phase. Otherwise, it is in a Sm$C$ phase.

After this prologue, we are now in the position to write down our model. Over all, the total elastic energy density $f$ of a smectic elastomer membrane will be of the form
\begin{align}
\label{totalEn}
f = f_{\text{iso}} + f_{\text{tilt}} + f_{\text{coupl}} + f_{\text{bend}}  \, .
\end{align}
In the following, we will for briefness often refer to energy densities somewhat loosely as energies. $f_{\text{iso}}$ is the well known stretching energy of isotropic polymerized membranes~\cite{paczuski_kardar_nelson_1988}. In terms of the metric tensor, it can be formulated as
\begin{align}
\label{isoEn}
f_{\text{iso}} = t \tr \tens{g} +  \textstyle{\frac{1}{2}} B \tr^2 \tens{g}  + \mu \tr \tens{\hat{g}}^2 \, ,
\end{align}
where $\hat{g}_{ab}=g_{ab}-\frac{1}{2}\delta_{ab} g_{cc}$ is the traceless variant of the metric tensor. $B$ and $\mu$ are, respectively, the bulk and shear moduli of the membrane. $t$ is a tunable parameter. In mean field theory, $f_{\text{iso}}$ predicts a second order transition from a flat to crumpled phase when $t$ changes sign from positive to negative. Real samples of smectic elastomers are essentially incompressible. To strictly enforce incompressibility, we had to use a term $\textstyle{\frac{1}{2}} B ( \det \tens{g} - 1)^2$, which fixes the membrane volume for $B \to \infty$, rather than $\textstyle{\frac{1}{2}} B \tr^2 \tens{g}$, which enforces incompressibility only at small but not at large strains. The more general term, however, would add algebraic complexity to our model without that it would change the results of our Landau-type theory qualitatively. In the equilibrium Sm$A$-flat phase of our smectic membrane, the director prefers to be parallel to the layer normal, and there are energy costs associated with deviations from this equilibrium, which are proportional to $\sin^2 \Theta$ and $\sin^4 \Theta$ etc., where $\Theta$ is the angle between the $\vec{N}$ and $\vec{n}$. This leads to the tilt energy
\begin{align}
\label{Htilt}
f_{\text{tilt}} &= \textstyle{\frac{1}{2}} \, r \, \tilde{c}_a^2 +\textstyle{\frac{1}{4}} \, v \, (\tilde{c}_a^2)^2 ,
\end{align}
with an adjustable parameter $r$. In mean field theory, $f_{\text{tilt}}$ predicts a Sm$A$ phase, where the c-director vanishes, for $r>0$ and a Sm$C$ phase, where the director has a component in the smectic plane, for $r<0$. $f_{\text{coupl}}$ is the coupling energy between the elastic and the liquid-crystalline degrees of freedom. When keeping only the lowest order terms permitted by symmetry, it is given by
\begin{align}
\label{Hcoupl}
f_{\text{coupl}} &= - \lambda_1 \, g_{aa} \tilde{c}_b^2 - \lambda_2 \, \tilde{c}_a   \, \hat{g}_{ab}  \, \tilde{c}_b \, ,
\end{align}
where $\lambda_1$ and $\lambda_2$ are coupling constants which we assume to be positive so that the coupling favors alignment of the c-director and the principle axes of the metric tensor. Finally, 
\begin{align}
\label{Hbend}
f_{\text{bend}} &= \textstyle{\frac{1}{2}} \, K \, \big(\partial_a^2 \vec{R} \big)^2
\end{align}
is a bending energy with a bending modulus $K$. In what follows, we can often disregard bending terms because they are of higher order in derivatives than the other contributions to the total elastic energy. At certain instances, however, namely when we deal with soft elasticity, bending terms will be important to ensure mechanical stability.

In bulk elastomers, the bulk and the shear moduli are typically of the order of $10^{9}$ Pa and $10^{6}$ Pa, respectively. The coefficient of the fourth order term in the tilt energy is of order $10^6$ Pa in smectic elastomers~\cite{brehmer&Co_1996} as it is in conventional smectics~\cite{ArcherDie2005}. We assume, that the orders of magnitude of corresponding quantities in our model follow the same hierarchy. Moreover, we assume that the coupling constants $\lambda_1$ and $\lambda_2$ are considerably smaller than the other elastic constants. Thus, our hierarchy of magnitudes is $B \gg \mu \sim v \gg \lambda_1\sim \lambda_2$. 

\section{Mean-Field Phase Diagram}
\label{meanFieldPhaseDiagram}
As mentioned above, the stretching energy of $f_{\text{iso}}$ of isotropic tethered membranes predicts a crumpling transition, and the tilt energy $f_{\text{tilt}}$ of smectic liquid crystals predicts a transition between Sm$A$ and Sm$C$. Through the coupling energy $f_{\text{coupl}}$, there is an interplay of conformational order and liquid crystalline order. By simply combining the phase characteristics of conventional smectics and conventional polymerized membranes, we expect that this interplay leads to the following phases: Sm$A$--flat, Sm$C$--flat, Sm$A$--crumpled, Sm$C$--crumpled and Sm$C$--tubule. Naively, one might also expect a Sm$A$--tubule phase. Such a phase, however, does not occur because any anisotropy in $\tens{g}$ acts like a temperature shift leading to  a non-vanish equilibrium c-director.
\begin{figure}
\centerline{\includegraphics[width=7.0cm]{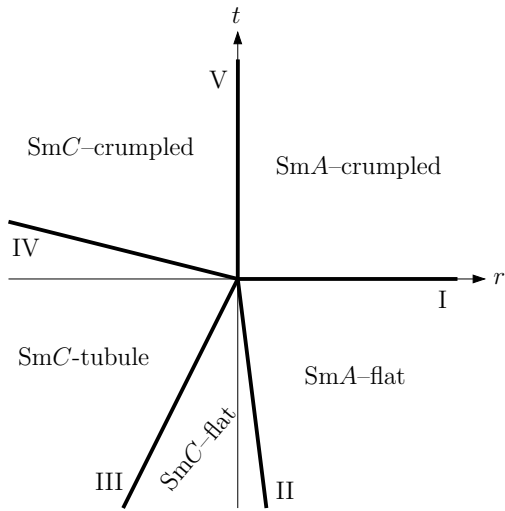}}
\caption{Schematic phase diagram of a smectic elastomer membrane in mean field theory. There are five phases, viz.\ Sm$A$--flat, Sm$C$--flat, Sm$A$--crumpled, Sm$C$--crumpled and Sm$C$--tubule separated by second order phase transitions (solid lines). The five second order lines meet at the origin, which, therefore, is a pentacritical point.}
\label{fig:phaseDiagram}
\end{figure}

To study the phase diagram in detail, we minimize the total elastic energy~(\ref{totalEn}) over the metric tensor and the c-director for given $t$ and $r$. To simplify this minimization, we choose our coordinates in reference space such that the metric tensor is diagonal,
\begin{align}
\label{gParametrization}
\tens{g} = \left(
\begin{array}{cc}
g_1 &  0 \\
0 & g_2
\end{array}
\right) ,
\end{align}
where $g_1$ and $g_2$, which are non-negative, are the two eigenvalues of $\tens{g}$. For the c-director, we employ the parametrization~\cite{footnote_cParametrization}
\begin{align}
\label{cParametrization}
\brm{\tilde{c}} = (\tilde{c}_1, \tilde{c}_2) = (S, 0)  \, .
\end{align}
Throughout this section we can omit the bending energy~(\ref{Hbend}); the higher-derivative bending terms do not influence the mean-field phase diagram because the equilibrium metric tensor is certainly uniform as is the equilibrium c-director.

Before embarking on the actual minimization procedure, we would like to comment on the connection between our model and $\Phi^4$-models. Introducing $3\times1$ matrices $\underline{\xi} = (\xi_1, \xi_2, \xi_3) = (g_1, g_2, S^2)$ and $\underline{t} = (t, t, r/2)$ and the symmetric $3\times 3$ matrix
\begin{align}
\label{Cdef}
\tens{C} = \left(
\begin{array}{ccc}
B + \mu &  B - \mu & - (\lambda_1 + \lambda_2/2) \\
B - \mu & B + \mu &  - (\lambda_1 -  \lambda_2/2) \\
- (\lambda_1 + \lambda_2/2) & - (\lambda_1 - \lambda_2/2) & v/2
\end{array}
\right) ,
\end{align}
our model elastic energy can be written as
\begin{align}
f = \underline{t} \cdot \underline{\xi} + \textstyle{\frac{1}{2}} \, \underline{\xi} \cdot \tens{C} \, \underline{\xi}\, .
\end{align}
This is the generic form of all generalized $\Phi^4$-models,  the most prominent example of which is the anisotropic antiferromagnet with two competing order parameters~\cite{antiferromagnet}. For the phase behavior of any of these models it is crucial wether the coupling matrix $\tens{C}$ is positive definite or not. If $\tens{C}$ is positive definite, all transitions in the mean-field phase diagram are continuous. Otherwise, one can have first order transitions. In the antiferromagnet, for example, one has four second order lines meeting in a tetracritical point if $\tens{C}$ is positive whereas one has two second order and one first order line meeting in a bicritical point if this is not the case. We will see shortly that $\tens{C}$ is positive definite in our model and thus all transitions in our mean-field phase diagram are continuous. Recalling from the beginning of this section that we anticipate five phases, we therefore expect to find five second order lines meeting in a pentacritical point.

For the actual minimization, we find it most convenient to write the total elastic energy in the following form,
\begin{subequations}
\label{fRecast}
\begin{align}
\label{fConvenient}
f &=  f^{(1)} + f^{(2)}, 
\end{align} 
where
\begin{align}
f^{(1)} & = \textstyle{\frac{1}{2}} B \left[ \xi_1 + \xi_2 - \gamma_1 - \gamma_2 \right]^2 
\nonumber \\
&+ \textstyle{\frac{1}{2}} \mu \left[ \xi_1 - \xi_2 - \gamma_1 + \gamma_2 \right]^2 \, ,
\\
f^{(2)} & = \textstyle{\frac{1}{4}} v_R \left[  \xi_3 + r_R/v_R \right]^2 \, .
\end{align}
\end{subequations}
Here, we dropped inconsequential constant terms. $\gamma_1$ and $\gamma_2$ are abbreviations for
\begin{subequations}
\begin{align}
\gamma_1 & = \frac{\alpha + \beta}{2} \, \xi_3 - \frac{t}{2\, B} \, ,
\\
\label{gamma2Def}
\gamma_2 & = \frac{\alpha - \beta}{2} \, \xi_3 - \frac{t}{2\, B} \, ,
\end{align}
\end{subequations}
where $\alpha = \lambda_1/B$ and $\beta = \lambda_2/(2\mu)$. Note that $\gamma_1 \geq \gamma_2$. Note also that with our assumed hierarchy of magnitudes, $\beta \gg \alpha$. $r_R$ and $v_R$ are renormalized versions of the elastic constants of the tilt energy,
\begin{subequations}
\begin{align}
r_R & =r + 2\,  \frac{\lambda_1}{B} \, t \, ,
\\
v_R & = v - 2\,  \frac{\lambda_1^2}{B} -  \frac{\lambda_2^2}{2\, \mu} \, .
\end{align}
\end{subequations}
Equation~(\ref{fRecast}) shows the total elastic energy in its diagonalized form. From it, we can read off the eigenvalues of $\tens{C}$, namely $B$, $\mu$ and $\frac{1}{2} v_R$. Knowing that $B$ and $\mu$ are positive, we conclude that $\tens{C}$ is indeed positive definite if $v_R >0$, which is the case for our assumed hierarchy of magnitudes. 

The energy $f^{(1)}$ has a simple geometrical interpretation that provides for intuitive guidance in the minimization process. When viewed in the $(\xi_1, \xi_2)$ plane, the contours of constant $f^{(1)}$ for fixed $\gamma_1$ and $\gamma_2$ are ellipses centered about the point $(\gamma_1, \gamma_2)$ with their short axis along $1/\sqrt{2} \,  (1,1)$ and their long axis along $1/\sqrt{2} \, (1,-1)$. Due to $\gamma_1 \geq \gamma_2$, there are 3 qualitatively different cases: (i) $\gamma_1 , \gamma_2 > 0$, (ii) $\gamma_1 >0, \gamma_2 <0$, and (iii) $\gamma_1 , \gamma_2 < 0$. In case (i), $f^{(1)}$ is minimized by $\xi_1^0 = \gamma_1$ and $\xi_2^0 = \gamma_2$, corresponding to the Sm$A$--flat or Sm$C$--flat phase. In case (ii), it is minimized by some $\xi_1^0>0$ and $\xi_2^0 = 0$, corresponding to the Sm$C$--tubule phase.  In case (iii), $f^{(1)}$ is minimal for $\xi_1^0 = \xi_2^0 = 0$ corresponding to the Sm$A$--crumpled or Sm$C$--crumpled phase.

Having discussed the possible equilibrium values qualitatively, we now turn to their actual calculation. The $\xi_\nu$, $\nu \in I \equiv \{1, 2, 3\}$, are non-negative, i.e., we have to minimize $f$ over the non-negative octant of $(\xi_1, \xi_2, \xi_3)$-space. This can be done straightforwardly by taking the derivatives of $f$, Eq.~(\ref{fRecast}), with respect to $\xi_\nu$ and then setting $\partial f/\partial \xi_{\nu^\prime} = 0$, $\nu^\prime \in I^\prime$,  for any subset $I^\prime$ of $I$ and setting $\xi_{\nu^{\prime\prime}} = 0$ for the corresponding complement,  $\nu^{\prime\prime} \in I^{\prime \prime} \equiv I/I^\prime$. This way we obtain sets of linear equations which are then solved subject to the condition $\xi_{\nu^\prime} > 0$. This procedure leads to the following phases as described by the following equilibrium values $\xi_\nu^0$:
\begin{itemize}
\item
Sm$A$--crumpled phase
\begin{align}
\xi_1^0 = \xi_2^0 = \xi_3^0 = 0 \, ,
\end{align}
\item
Sm$A$--flat phase
\begin{subequations}
\begin{align}
\xi_1^0 &= \xi_2^0 = | t| /(2B) \, ,
\\
\xi_3^0 &= 0 \, ,
\end{align}
\end{subequations}
\item
Sm$C$--crumpled phase
\begin{subequations}
\begin{align}
\xi_1^0 &= \xi_2^0 = 0 \, ,
\\
\xi_3^0 &= |r|/v \, ,
\end{align}
\end{subequations}
\item
Sm$C$--flat phase
\begin{subequations}
\begin{align}
\xi_1^0 &= \left|  \frac{\alpha + \beta}{2} \, \frac{r_R}{v_R} + \frac{t}{2B} \right| ,
\\
\xi_2^0 & = \left|  \frac{\alpha - \beta}{2} \, \frac{r_R}{v_R} + \frac{t}{2B} \right|  ,
\\
\xi_3^0 &= |r_R|/v_R \, ,
\end{align}
\end{subequations}
\item
Sm$C$--tubule phase
\begin{subequations}
\begin{align}
\xi_1^0 &= \frac{\lambda_1 + \lambda_2/2}{\bar{v}_R (B +\mu)} \left|  r + \frac{v t}{\lambda_1 + \lambda_2/2} \right| ,
\\
\xi_2^0 & = 0 \, ,
\\
\xi_3^0 &= |\bar{r}_R|/\bar{v}_R  ,
\end{align}
\end{subequations}
\end{itemize}
where
\begin{subequations}
\begin{align}
\bar{r}_R & =r + 2\,  \frac{\lambda_1 + \lambda_2/2}{B+\mu} \, t \, ,
\\
\bar{v}_R & = v - 2\,   \frac{(\lambda_1 + \lambda_2/2)^2}{B+\mu} \, .
\end{align}
\end{subequations}

The remaining task for assessing the mean-field phase diagram is to determine the boundaries between these phases. This can be done economically by setting $\partial f/\partial \xi_{\nu} = 0$ and $\xi_{\nu^\prime} = 0$, where, as above, $\nu \in I$ and $\nu^\prime \in I^\prime$ for any subset $I^\prime$ of $I$. Solving the so-obtained sets of linear equations results in the following second order phase transition lines which we label as shown in Fig.~\ref{fig:phaseDiagram} by capital roman numbers:
\begin{itemize}
\item
Line I
\begin{align}
t = 0\, , \quad r>0 \, ,
\end{align}
\item
Line II
\begin{align}
t = - \frac{B}{2\, \lambda_1} \, r \, , \quad r > 0 \, ,
\end{align}
\item
Line III
\begin{align}
t = \frac{B(\beta  - \alpha)}{v_R - 2 \lambda_1 (\beta - \alpha)} \, r \, ,  \quad r < 0 \, ,
\end{align}
\item
Line IV
\begin{align}
t = - \frac{\lambda_1 + \lambda_2/2}{v} \, r \, , \quad r < 0 \, ,
\end{align}
\item
Line V
\begin{align}
r = 0 \, , \quad t > 0 \,  .
\end{align}
\end{itemize}
All five second order lines meet at the origin of $(r, t)$-space. Thus, this origin is a pentacritical point. Note that line III approaches line II for $v_R$ approaching zero, i.e., the area in phase space occupied by the Sm$C$--flat phase becomes vanishingly small in this limit. 

Having discussed the phase diagram in mean-field theory, it is a legitimate and interesting question to ask inasmuch the mean-field phase diagram will be modified by the effects of fluctuations, self-avoidance and random stresses. A definite answer to this question requires renormalization group analyses and is  beyond the scope of this paper. Given the previous work on isotropic and anisotropic polymerized membranes, however, one can speculate what might happen. Paczuski, Kardar, and Nelson~\cite{paczuski_kardar_nelson_1988} showed that the crumpled-to-flat transition in isotropic membranes is driven first order by fluctuations for embedding dimensions $d<d_c = 219$. Because there is no in-plane anisotropy in the Sm$A$--flat and Sm$A$--crumpled phases of smectic elastomer membranes, the transition between these phases might also turn out to be a fluctuation-driven first order transition. In their work on permanently anisotropic membranes, Radzihovsky and Toner~\cite{radzihovsky_toner_anisoMem} found that the crumpled-to-tubule transition remains second order for all $d$ and they concluded that fluctuations do not change the topology of the phase diagram. This might indicate that fluctuations will not modify qualitatively the locus and the order of the transition between the Sm$C$--flat and the Sm$C$--tubule and the Sm$C$--tubule and the Sm$C$--crumpled phases.

\section{Elasticity Of The Sm$C$--Flat Phase}
\label{elasticityOfTheFlatPhase}
As mentioned in the introduction, the elasticity of the Sm$C$--flat and the Sm$C$--tubule phases is, among that of the various phases featured in the phase diagram, Fig.~\ref{fig:phaseDiagram}, the most interesting. In these phases, rotational symmetry in reference space is spontaneously broken and thus, due to the Goldstone theorem, the membrane has zero-energy long-wavelength modes. Here, we study these Goldstone modes in some detail for the Sm$C$--flat phase.

In Sec.~\ref{meanFieldPhaseDiagram} we learned that the Sm$C$--flat phase is characterized by an equilibrium metric tensor with two different and non-vanishing eigenvalues $g_1^0 > g_2^0 > 0$ and a non-vanishing equilibrium c-director, i.e., $S^0 \neq 0$. To keep our discussion of the elasticity of the Sm$C$--flat phase as simple as possible, we choose our bases for the reference and target spaces such that $\hat{e}_x \equiv \hat{e}_1$ and $\hat{e}_y \equiv \hat{e}_2$ are along the eigenvectors pertaining to $g_1^0$ and $g_2^0$, respectively. With these choices, the equilibrium or reference conformation of the Sm$C$--flat membrane is characterized by
\begin{align}
\vec{R}^0 (\brm{x}) = \zeta_1 \, x \, \hat{e}_x +  \zeta_2 \, y \, \hat{e}_y \, ,
\end{align}
 where $\zeta_1 = \sqrt{g_1^0}$ and $\zeta_2 = \sqrt{g_2^0}$. To describe deviation from this equilibrium, it is useful to employ a two-dimensional elastic displacement field $\brm{u}(\brm{x})$ with components $u_x (\brm{x})$ and $u_y (\brm{x})$ and a one-dimensional out-of-plane undulation (height) field $h (\brm{x})$:
\begin{align}
\vec{R} (\brm{x}) = [ \zeta_1 \, x  + u_x (\brm{x}) ] \hat{e}_x +  [ \zeta_2 \, y + u_y (\brm{x}) ] \hat{e}_y  +  h (\brm{x}) \, \hat{e}_z \, ,
\end{align}
where $\hat{e}_z = \hat{e}_3$. With this parametrization, the metric tensor reads
\begin{align}
\label{gParametrizationSmCflat}
\tens{g} = \left(
\begin{array}{cc}
\zeta_1^2 + 2 u_{xx} &  2 u_{xy} \\
2 u_{xy} &\zeta_2^2 + 2 u_{yy}
\end{array}
\right) ,
\end{align}
with the components of the strain tensor, Eq.~(\ref{defStrainTensor}), given by
\begin{subequations}
\label{SmCflatStrain}
\begin{align}
u_{xx} & = \textstyle{\frac{1}{2}} \left\{ 2 \zeta_1  \partial_x u_x + \partial_x \brm{u} \cdot  \partial_x \brm{u} + (\partial_x h)^2 \right\}  ,
\\
u_{xy} & = \textstyle{\frac{1}{2}} \big\{ (\zeta_1 + \partial_x u_x) \partial_y u_x  +(\zeta_2 + \partial_y u_y) \partial_x u_y 
\nonumber \\
&+  \partial_x h\partial_y h \big\}  ,
\\
u_{yy} & = \textstyle{\frac{1}{2}} \left\{ 2 \zeta_2  \partial_y u_y + \partial_y \brm{u} \cdot  \partial_y \brm{u} + (\partial_y h)^2 \right\}  .
\end{align}
\end{subequations}
For the c-director, we use parametrization
\begin{align}
\label{cParametrizationSmCflat}
\brm{\tilde{c}} = (\sigma + \delta \tilde{c}_x , \delta \tilde{c}_y)  \, ,
\end{align}
with $\sigma= S^0$, and with $\delta \tilde{c}_x$ and $\delta \tilde{c}_y$ describing longitudinal and transversal deviations from equilibrium, respectively. 

Now, we expand the elastic energy about the Sm$C$--flat ground state by inserting the metric tensor and the c-director as parametrized, respectively, by Eqs.~(\ref{gParametrizationSmCflat}) and (\ref{cParametrizationSmCflat}) into the elastic energy~(\ref{totalEn}). To guarantee that $\zeta_1$, $\zeta_2$ and $\sigma$ describe the true Sm$C$--flat ground state, they have to satisfy equations of state determined by the condition that terms in the deviation $\delta f$ of the elastic energy from its equilibrium value in the Sm$C$--flat phase that are linear in $u_{xx}$, $u_{yy}$ and $\delta \tilde{c}_x$ must vanish,
\begin{subequations}
\label{eosSmCflat}
\begin{align}
& t + B (\zeta_1^2 + \zeta_2^2) + \mu   (\zeta_1^2 - \zeta_2^2) - (\lambda_1 + \lambda_2/2) \sigma^2 = 0 \, ,
\\
&t + B (\zeta_1^2 + \zeta_2^2) - \mu   (\zeta_1^2 - \zeta_2^2) - (\lambda_1 - \lambda_2/2) \sigma^2 = 0 \, ,
\\
& \left[  r + v \sigma^2 - 2 \lambda_1 (\zeta_1^2 + \zeta_2^2) - \lambda_2 (\zeta_1^2 - \zeta_2^2) \right]  \sigma = 0 \, .
\end{align}
\end{subequations}
Given that these equations of state are satisfied, we obtain
\begin{align}
\label{deltaFSmCflat}
\delta f &= v \sigma^2  (\delta \tilde{c}_x)^2 - 2 [(2\lambda_1 + \lambda_2) u_{xx} + (2\lambda_1 - \lambda_2) u_{yy}] \delta \tilde{c}_x
\nonumber \\
& + 8 \mu \left\{ u_{xy} - \frac{\zeta_1^2 - \zeta_2^2}{2 \, \sigma} \, \delta \tilde{c}_y \right\}^2 
\nonumber \\
&+ 2 ( B + \mu) [u_{xx}^2 + u_{yy}^2] + 4  (B - \mu) u_{xx} u_{yy}
\end{align}
to harmonic order in the strains and the $\delta \tilde{c}_a$. In the spirit of Landau theory, due to the term  $v \sigma^2  (\delta \tilde{c}_x)^2$, the longitudinal deviation $\delta \tilde{c}_x$ is a massive variable and, therefore, the relaxation of $\delta \tilde{c}_x$ cannot be the origin of the anticipated softness of the Sm$C$--flat phase. Thus, we integrate this massive variable out, i.e.\ we replace it by its equilibrium value in the presence of strain, which leads to  
\begin{align}
\label{deltaFSmCflatNoMassive}
\delta f &=   8 \mu \left\{ u_{xy} - \frac{\zeta_1^2 - \zeta_2^2}{2 \, \sigma} \, \delta \tilde{c}_y \right\}^2 
\nonumber \\
&+ 2 \left[   B + \mu - 2 \frac{(\lambda_1 + \lambda_2/2)^2}{v}\right] u_{xx}^2
\nonumber \\
&+ 2 \left[   B + \mu - 2 \frac{(\lambda_1 - \lambda_2/2)^2}{v}\right] u_{yy}^2 
\nonumber \\
&+ 4  \left[   B - \mu - 2 \frac{(\lambda_1 + \lambda_2/2)(\lambda_1 - \lambda_2/2)}{v}\right] u_{xx} u_{yy} \, .
\end{align}
This equation shows how $\delta \tilde{c}_y$ can relax locally to eliminate the dependence of the elastic energy on $u_{xy}$. In other words, a smectic elastomer membrane in the Sm$C$--flat phase is soft with respect to shear in the plane of the membrane.

The strain $\tens{u}$ describes distortions relative to the new Sm$C$--flat reference state measured in the coordinates of the old Sm$A$--flat reference state. However, it is more intuitive and more customary to use the natural coordinates $x^\prime = R^0_x = \zeta_1x$ and $y^\prime = R^0_y = \zeta_2 y$ of the new state. Expressed in terms of the strain $\tens{u}^\prime$, whose components are related to those of $\tens{u}$ by $u_{xx} = \zeta_1^2 u_{xx}^\prime$, $u_{xy} = \zeta_1\zeta_2 u_{xy}^\prime$ and $u_{yy} = \zeta_2^2 u_{yy}^\prime$, the elastic energy can be written as
\begin{align}
\label{deltaFSmCflatFinal}
\delta f &=  \textstyle{\frac{1}{2}} C_{xxxx} (u_{xx}^\prime)^2 + \textstyle{\frac{1}{2}} C_{yyyy} (u_{yy}^\prime)^2 + C_{xxyy} u_{xx}^\prime u_{yy}^\prime
\nonumber \\
&+ \textstyle{\frac{1}{2}} \kappa_{xx} (\partial_x^{\prime2} h )^2 + \textstyle{\frac{1}{2}} \kappa_{yy} (\partial_y^{\prime2} h )^2 + \kappa_{xy} (\partial_y^{\prime2} h ) (\partial_x^{\prime2} h )
\nonumber \\
&+ \textstyle{\frac{1}{2}} K_y (\partial_y^{\prime2} u_x )^2 + \textstyle{\frac{1}{2}} K_x (\partial_x^{\prime2} u_y )^2  ,
\end{align} 
where $\partial_x^\prime$ and $\partial_y^\prime$ are abbreviations for $\partial/ (\partial x^\prime)$ and $\partial/ (\partial y^\prime)$, respectively. The elastic constants of the stretching terms are given by
\begin{subequations}
\label{SmCflatStretchingConstants}
\begin{align}
C_{xxxx} & = 4 \zeta_1^4  \left[   B + \mu - 2 \frac{(\lambda_1 + \lambda_2/2)^2}{v}\right]   ,
\\
C_{yyyy} & = 4 \zeta_2^4 \left[   B + \mu - 2 \frac{(\lambda_1 - \lambda_2/2)^2}{v}\right]  ,
\\
C_{xxyy} & = 4 \zeta_1^2 \zeta_2^2 \left[   B - \mu - 2 \frac{(\lambda_1 + \lambda_2/2)(\lambda_1 - \lambda_2/2)}{v}\right] .
\end{align}
\end{subequations}
As already pointed out above, there is no term of the type $C_{xyxy} (u_{xy}^\prime)^2$ because the membrane shear modulus $C_{xyxy}$ vanishes as a result of the broken rotational symmetry of the Sm$C$--flat phase. Due to this soft elasticity, we added in Eq.~(\ref{deltaFSmCflatFinal}) bending terms stemming from the bending energy~(\ref{Hbend}) to ensure mechanical stability. As can be easily checked, the bending constants are given by $\kappa_{xx} = \zeta_1^4 K$, $\kappa_{yy} = \zeta_2^4 K$ and so on.

Our final elastic energy~(\ref{deltaFSmCflatFinal}) is identical in form to the harmonic elastic energy of two-dimensional nematic elastomer membranes in their flat phase. The only differences lie in the values of the elastic constants. Nematic elastomer membranes, including their generalizations to $D$-dimensional nematic--flat membranes in $d$-dimensional embedding space, have been studied in detail in Ref.~\cite{Xing&Co_fluctNemMem_2003}. For example, Ref.~\cite{Xing&Co_fluctNemMem_2003} contains a detailed analysis of correlations and fluctuations in mean-field theory. These results can be transcribed directly to Sm$C$--flat membranes with the only differences residing in the specific values of the elastic constants. To safe space, we refrain here from further commenting on correlations, fluctuations and refer directly to Ref.~\cite{Xing&Co_fluctNemMem_2003}.

Generically, fluctuation effects are strong in soft phases. Fluctuations drive elastic nonlinearities, which are often negligible in systems without soft elasticity, to qualitatively modify the elasticity through a Grinstein-Pelcovits-type renormalization~\cite{grinstein_pelcovits_81_82}. As a consequence of this renormalization, the elasticity becomes anomalous with length-scale dependent elastic constants (in the form of power laws with universal scaling exponents or logarithmic corrections, depending on dimensionality) and universal Poisson ratios. Reference~\cite{Xing&Co_fluctNemMem_2003} presents a renormalization group study of fluctuation effects in the flat phase of nematic elastomer membranes. Because this phase and the Sm$C$--flat phase share the same macroscopic symmetries, we expect their anomalous elasticity to be governed by the same universal quantities, for which we refer to~\cite{Xing&Co_fluctNemMem_2003}. As far as self-avoidence is concerned, it is known that this effect is irrelevant in physical dimensions for flat permanently anisotropic polymerized membranes~\cite{radzihovsky_toner_anisoMem}. We expect this irrelevance also to hold for flat nematic and Sm$C$ elastomer membranes.

\section{Elasticity Of The Sm$C$--Tubule Phase}
\label{elasticityOfTheTubulePhase}
As in the Sm$C$--flat phase, rotational symmetry is spontaneously broken in the Sm$C$--tubule phase and, therefore, also the Sm$C$--tubule phase should be expected on grounds of the Goldstone theorem to exhibit soft elasticity. Here, we will study the elasticity of the Sm$C$--tubule phase in some detail.

First, let us recall that the Sm$C$--tubule phase is characterized by an equilibrium metric tensor with one vanishing and one positive eigenvalue, $g_1^0 >  0,  g_2^0 = 0$ and a non-vanishing equilibrium c-director, $S^0 \neq 0$. Choosing our basis so that $\hat{e}_x$ is along the eigenvector associated with $g_1^0 $, the reference conformation of the Sm$C$--tubule phase is characterized by
\begin{align}
\vec{R}^0 (\brm{x}) = \zeta \, x \, \hat{e}_x  \, ,
\end{align}
where $\zeta = \sqrt{g_1^0}$. To describe distortions, we here employ a one-dimensional elastic displacement field $u (\brm{x})$ and a two-dimensional height field $\brm{h}(\brm{x})$ with components $h_y (\brm{x})$ and $h_y (\brm{x})$. With this parametrization, we have
\begin{align}
\vec{R} (\brm{x}) = [ \zeta_1 \, x  + u (\brm{x}) ] \hat{e}_x + h_y (\brm{x})  \, \hat{e}_y  +  h_z  (\brm{x}) \, \hat{e}_z 
\end{align}
for the target space coordinate of the mass point $\brm{x}$ after distortion and 
\begin{align}
\label{gParametrizationSmCtubule}
\tens{g} = \left(
\begin{array}{cc}
\zeta^2 + 2 u_{xx} &  2 u_{xy} \\
2 u_{xy} &  2 u_{yy}
\end{array}
\right) 
\end{align}
for the corresponding metric tensor. The components of the strain tensor featured in Eq.~(\ref{gParametrizationSmCtubule}) read
\begin{subequations}
\label{SmCtubuleStrain}
\begin{align}
u_{xx} & = \textstyle{\frac{1}{2}} \left\{ 2 \zeta  \partial_x u + (\partial_x u)^2 + \partial_x \brm{h} \cdot  \partial_x \brm{h} \right\}  ,
\\
u_{xy} & = \textstyle{\frac{1}{2}} \left\{ (\zeta + \partial_x u) \partial_y u   +  \partial_x \brm{h} \cdot  \partial_y \brm{h} \right\}  ,
\\
u_{yy} & = \textstyle{\frac{1}{2}} \left\{  (\partial_y u)^2 +  \partial_y \brm{h} \cdot  \partial_y \brm{h} \right\}  .
\end{align}
\end{subequations}
For the c-director, we can use the same parametrization as for the Sm$C$--flat phase, see Eq.~(\ref{cParametrizationSmCflat}). 

Next, we substitute Eqs.~(\ref{gParametrizationSmCtubule}) and (\ref{cParametrizationSmCflat}) into the elastic energy~(\ref{totalEn}) and expand to harmonic order in the strains and $\delta \tilde{c}_a$. Because $\zeta$ and $\sigma$ characterize the equilibrium values of the metric tensor and the c-director, they satisfy equations of state,
\begin{subequations}
\label{eosSmCtubule}
\begin{align}
& t + (B  + \mu) \zeta^2  - (\lambda_1 + \lambda_2/2) \sigma^2 = 0 \, ,
\\
& \left[  r + v \sigma^2 - 2 (\lambda_1 + \lambda_2/2) \zeta^2)  \right]  \sigma = 0 \, .
\end{align}
\end{subequations}
such that there are no terms linear in $u_{xx}$ or $\delta \tilde{c}_x$ in the expanded elastic energy: \begin{align}
\label{deltaFSmCtubule}
\delta f &= 2(\lambda_2 \sigma^2 - 2 \mu \zeta^2) \, u_{yy} + 2 ( B + \mu) [u_{xx}^2 + u_{yy}^2]
\nonumber \\
& + 4  (B - \mu) \, u_{xx} u_{yy} + 8 \mu \, u_{xy}^2 +v \sigma^2  (\delta \tilde{c}_x)^2 + \lambda_2 \zeta^2 (\delta \tilde{c}_y)^2
\nonumber \\
& - 4  (\lambda_1 + \lambda_2/2) \sigma \, u_{xx} \delta \tilde{c}_x - 4  (\lambda_1 - \lambda_2/2) \sigma \, u_{yy} \delta \tilde{c}_x 
\nonumber \\
&- 4 \lambda_2 \sigma \, u_{xy} \delta \tilde{c}_y \, .
\end{align}
Comparing Eq.~(\ref{deltaFSmCtubule}) to Eq.~(\ref{deltaFSmCflat}), we note the following qualitative difference: in the case of the Sm$C$--flat phase, the terms depending on $u_{xy}$ and $\delta \tilde{c}_y$ combine to form a complete square; in the case of the Sm$C$--tubule phase they do not. Hence, in the latter case the relaxation of $\delta \tilde{c}_y$  cannot eliminate the dependence of the elastic energy on $u_{xy}$ entirely. To determine what kinds of deformation are actually soft in this phase, we now switch from the strains to the elastic displacement and hight fields:
\begin{align}
\label{deltaFSmCtubuleUH}
\delta f &= u \sigma^2  (\delta \tilde{c}_x)^2 - 4  (\lambda_1 + \lambda_2/2) \zeta \sigma \, \partial_x u \, \delta \tilde{c}_x
\nonumber \\
& + \lambda_2 \zeta^2 \left\{   \delta \tilde{c}_y -  \frac{\sigma}{\zeta} \partial_y u\right\}^2 +  2 ( B + \mu) \zeta (\partial_x u)^2 
\nonumber \\
&  + (\lambda_2 \sigma^2 - 2 \mu \zeta^2)\,  \partial_y \brm{h} \cdot  \partial_y \brm{h} \, ,
\end{align}
where we discarded all terms of higher than harmonic order. Equation~(\ref{deltaFSmCtubuleUH}) makes it transparent that $\delta \tilde{c}_y$ can relax locally to $\delta \tilde{c}_y = (\sigma/\zeta) \partial_y u$ such that the dependence of the elastic energy on $\partial_y u$ is eliminated. Note that $\partial_y u$ is, up to constants, the linear part of the shear strain $u_{xy}$. Therefore, the Sm$C$--tubule phase exhibits soft elasticity with respect to this shear provided that it is small enough such that its nonlinear contributions can be neglected. As announced above, the origin of this softness is spontaneous breaking of the rotational symmetry of the initial Sm$A$--flat phase due to Sm$C$ ordering. Another observation that we make from Eq.~(\ref{deltaFSmCtubuleUH}) is that $\delta \tilde{c}_x$ is a massive variable, as it is in the Sm$C$--flat phase. Thus, we integrate it out, i.e., we replace it by its equilibrium value $\delta \tilde{c}_x= 2 (\lambda_1 + \lambda_2/2) \zeta/(v \sigma) \partial_x u$.  Another step that is worthwhile taking at this point is to switch from $x$ and $y$, which still pertain to the initial Sm$A$--flat phase, to the natural coordinates $x^\prime = R^0_x = \zeta x$ and $y^\prime = y$ of the Sm$C$--tubule phase. Eventually, we obtain
\begin{align}
\label{deltaFSmCtubuleFinal}
\delta f &=  \textstyle{\frac{1}{2}}  B_u  \, (\partial_x^\prime u)^2 +  \textstyle{\frac{1}{2}}  B_h \,  \partial_y^\prime \brm{h} \cdot  \partial_y^\prime \brm{h}
\nonumber \\
&+  \textstyle{\frac{1}{2}}  K_u  \, (\partial_y^{\prime 2} u)^2  +  \textstyle{\frac{1}{2}}  K_h \,  \partial_x^{\prime2} \brm{h} \cdot  \partial_x^{\prime2} \brm{h} \, ,
\end{align}
where $\partial_x^\prime = \partial/(\partial x^\prime)$, $\partial_y^\prime = \partial_y$, and where we added bending terms stemming from Eq.~(\ref{Hbend}) to ensure mechanical stability under soft deformations. The elastic constants of the stretching terms are given by
\begin{subequations}
\label{SmCtubuleStretchingConstants}
\begin{align}
B_u & = 4 \zeta^4  \left[   B + \mu - 2 \frac{(\lambda_1 + \lambda_2/2)^2}{v}\right]   ,
\\
B_h & = 2 \left[    \lambda_2\sigma^2 - 2 \zeta^2 \mu\right]  ,
\end{align}
\end{subequations}
and the bending moduli are given by $K_u = K$ and $K_h = \zeta^4 K$.

Equation~(\ref{deltaFSmCtubuleFinal}) is identical in form to the harmonic elastic energy of the nematic--tubule phase that has been studied extensively in Ref.~\cite{Xing_Radzihovsky_nemTubule_2005}. The only differences reside in the specifics of the elastic constants. Therefore, macroscopic properties of Sm$C$--tubule and nematic--tubule membranes are qualitatively the same in mean-field theory, at least as far as they can be captured by a model elastic energy in terms of elastic displacement and height fields only. This applies for example to the Gaussian correlation and fluctuations of the displacement and height fields. For details on these, we refer to Ref.~\cite{Xing_Radzihovsky_nemTubule_2005}. 

Fluctuations will presumably lead via a Grinstein-Pelcovits-type renormalization to anomalous elasticity of the Sm$C$--tubule phase. Because the nematic--tubule and the Sm$C$--tubule phases have the same macroscopic symmetries, the universal quantities characterizing the anomalous elasticity of the two phases are expected to be the same. To date, no renormalization group study of these universal quantities exist, although Ref.~\cite{Xing_Radzihovsky_nemTubule_2005} presents a minimal model that could be used as a vintage point for such a study. Self-avoidence is known to be relevant in physical dimensions for the tubule-phase in permanently anisotropic polymerized membranes~\cite{radzihovsky_toner_anisoMem}. We expect this relevance also for the tubule phases of nematic and Sm$C$ elastomer membranes.

\section{Concluding remarks}
\label{concludingRemarks}
In summary, we have developed a model for smectic elastomer membranes which includes elastic and liquid crystalline degrees of freedom. Based on our model, we determined the qualitative phase diagram of a elastomer membrane using mean-field theory. This phase diagram comprises five phases, viz.\ Sm$A$--flat, Sm$A$--crumpled, Sm$C$--flat, Sm$C$--crumpled and Sm$C$--tubule. Transition between adjacent phases are second order transitions. The harmonic elasticity of the Sm$C$--flat and Sm$C$--tubule phases is qualitatively the same (up to values of elastic constants) as that of the nematic--flat and the nematic--tubule phases, respectively, in nematic elastomer membranes. In particular, because they are all associated with a spontaneous breaking of in-plane rotational symmetry, these phases all exhibit soft elasticity, with the softness of the flat phases being qualitatively different from that of the tubule phases.

As far as future directions are concerned, it should be worthwhile to go beyond mean-field theory and to study the effects of non-linear elasticity and thermal fluctuation in renormalized field theory. Moreover, it should be interesting to proceed to a more realistic model by including self-avoidance and random stresses. Fluctuations, self-avoidance and random stresses will lead to qualitative modifications of at least some of our results and, therefore, understanding them will be one of our goals for future research.

As mentioned in the introduction, measurements of the elastic properties of thin films of smectic elastomers have been performed using a balloon geometry. In Ref.~\cite{StannariusZen2002} the authors find that for a chiral Sm$C^\ast$ elastomer film the balloon radius as a function of pressure deviates from the predictions of a simple phenomenological (Mooney-Rivlin) model and they hint that this deviation is related to soft elasticity. We hope, that our work motivates further experiments investigating the soft elasticity of smectic elastomer membranes in more detail. Moreover, we hope to encourage experiments on the phase behavior of smectic elastomer membranes that could be compared to our predictions for their phase diagram.

\acknowledgments
We thank H.-K.\ Janssen for a critical reading of the manuscript and helpful comments. We are particularly grateful to T.~C.~Lubensky for stimulating discussions and input at the early stages of this project.  



\begin{thebibliography}{}
\bibitem{footnote_cParametrization}
Strictly speaking, we should parametrize the c-director as $(\tilde{c}_1, \tilde{c}_2) = S (\cos \theta, \sin \theta)$ with some angle $\theta$. Then, however, minimization of $f$ with respect to $\theta$ reveals that in equilibrium $(\tilde{c}_1, \tilde{c}_2)$ lies along either one of the eigenvectors of $\tens{g}$. Without loss of generality we can choose it lie along the eigenvector pertaining to $g_1$ wich leads to the parametrization~(\ref{cParametrization}).
\end{thebibliography}
\end{document}